\documentclass[aps,prb,amsmath,amssymb,reprint,a4paper,showpacs, citeautoscript,longbibliography,floatfix]{revtex4-1}
\usepackage{mathptmx}
\usepackage[utf8]{inputenc} 
\usepackage{graphicx}
\usepackage{xspace}
\usepackage{epstopdf}
\usepackage[
,textwidth=17.5cm
,textheight=23.5cm
,verbose
,dvips
]{geometry}

\newcommand{\jap}{J.\ Appl.\ Phys.\xspace}

\begin{document}

\title{Luminescence of GaAs nanowires consisting of wurtzite and zincblende segments}
\author{U.\ Jahn}
\email{jahn@pdi-berlin.de}
\author{J.\ L\"ahnemann}
\author{C.\ Pf\"uller}
\author{O.\ Brandt}
\author{S.\ Breuer}
\author{B.\ Jenichen}
\author{M.\ Ramsteiner}
\author{L.\ Geelhaar}
\author{H.\ Riechert}
\affiliation{Paul-Drude-Institut f\"ur
Festk\"orperelektronik, Hausvogteiplatz 5--7, D-10117 Berlin,
Germany}

\begin{abstract}
GaAs nanowires (NWs) grown by molecular-beam epitaxy may contain segments of both the zincblende (ZB) and wurtzite (WZ) phases. Depending on the growth conditions, we find that optical emission of such NWs occurs either predominantly above or below the band gap energy of ZB GaAs ($E_g^\mathrm{ZB}$). This result is consistent with the assumption that the band gap energy of wurtzite GaAs ($E_g^\mathrm{WZ}$) is larger than $E_g^\mathrm{ZB}$ and that GaAs NWs with alternating ZB and WZ segments along the wire axis establish a type II band alignment, where electrons captured within the ZB segments recombine with holes of the neighboring WZ segments. Thus, the corresponding transition energy depends on the degree of confinement of the electrons, and transition energies exceeding $E_g^\mathrm{ZB}$ are possible for very thin ZB segments. At low temperatures, the incorporation of carbon acceptors plays a major role in determining the spectral profile as these can effectively bind holes in the ZB segments. From cathodoluminescence measurements of single GaAs NWs performed at room temperature, we deduce a lower bound of 55~meV for the difference $E_g^\mathrm{WZ} - E_g^\mathrm{ZB}$. 
\end{abstract}


\maketitle

\section{Introduction}

The growth and investigation of semiconductor nanowires (NWs) represent a quickly growing research field due to their potential application in electronic and optoelectronic devices. The equilibrium modification of bulk III-As semiconductors is the zincblende (ZB) phase. When grown in the NW form, however, these materials crystallize partially or even predominantly in the wurtzite (WZ) structure \cite{koguchi,mattila,tomioka}. In particular, GaAs NWs often exhibit a mixed crystal structure with ZB and WZ segments alternating along the NW axis \cite{soshnikov,spirkoska}. Since these phases differ from each other in their band gap, GaAs NWs represent heterostructures with complex optical properties, the interpretation of which is currently discussed controversially. In particular, the value of the band gap energy of WZ GaAs is under debate. On the one hand, it is reported that the band gap energy of WZ GaAs ($E_g^\mathrm{WZ}$) is larger by 30 to 50~meV than the one of ZB GaAs  ($E_g^\mathrm{ZB}$).\cite{murayama,zanolli,jancu,hoang1,hoang2,ihn}
On the other hand, results of theoretical and experimental investigations suggest that $E_g^\mathrm{WZ}$ is slightly smaller than $E_g^\mathrm{ZB}$ or on the same level. \cite{de,moewe,novikov,heiss,ketterer,ketterer1} 

There is a broad consensus that the WZ-ZB interface in GaAs exhibits a type II band alignment, where the conduction and valence band energy levels are higher in WZ than in ZB GaAs. Since the carrier diffusion length in GaAs is usually large (on the order of 1~$\mu$m), electrons are very efficiently captured within ZB segments. This results in a large probability to predominantly observe luminescence at energies lower than the band gap energies of either polytype, even if only a small number of ZB segments exists within the structure. Moreover, the presence of defect levels often favors the observation of luminescence at energies below that of the band gap, particularly for measurements at low temperatures ($T$). Thus, a clear interpretation of luminescence spectra of GaAs NWs requires more detailed investigations, where the most crucial question is the one about the actual value of $E_g^\mathrm{WZ}$. 

In this paper, we present experimental results on GaAs NWs containing both phases, where depending on the growth conditions luminescence is observed predominantly above or below $E_g^\mathrm{ZB}$. The observed luminescence spectra are in agreement with the assumption of a value of $E_g^\mathrm{WZ}$ larger than the one of $E_g^\mathrm{ZB}$, where electrons captured within the ZB segments recombine with holes of the neighboring WZ segments. Thus, the energy of the corresponding luminescence lines exceeds $E_g^\mathrm{ZB}$ for very thin ZB segments due to electron confinement.

\section{Experimental}

GaAs NWs were prepared by molecular-beam epitaxy (MBE) using Ga- and Au-induced vapor-liquid-solid (VLS) growth on phosphorous-doped on-axis ($\pm 1^\circ$) Si(111) substrates. The native oxide was intentionally left on the substrates, and a water desorption step was performed
under ultra high vacuum at 300$^\circ$C for 20~min. In the MBE chamber, the Si substrate was heated to 580$^\circ$C, a constant As$_4$ beam equivalent pressure of $1.2\times10^{-5}$~mbar was set, and GaAs nanowire growth started on opening of the Ga shutter. The total growth time was 1~h.
For sample \#1, the Ga supply was set to match a planar growth rate of 820~nm/h, which led to an atomic flux ratio of $F_{\text{As}}/F_\text{Ga}=1$. Stoichiometric conditions ($F_{\text{As}}/F_\text{Ga}=1$) were identified by the transition between the Ga-rich $(4\times2)$ and the As-rich $(2\times4)$ reconstruction for planar growth on GaAs(001) at 580$^\circ$C.\cite{daweritz}
In contrast, sample \#2 was grown using half the Ga flux ($F_{\text{As}}/F_\text{Ga}=2$) under otherwise identical growth conditions.
For both samples, the growth of an (Al,Ga)As shell with a nominal Al content of 10\% was initiated after 30~min by opening the Al shutter. A coverage of the GaAs NWs with an (Al,Ga)As shell was necessary to enhance the luminescence efficiency enabling optical investigations at moderate and low excitation densities \cite{breuer}. 

As a reference sample (\#Ref), bare GaAs NWs, which were known to crystallize predominantly in the WZ structure,\cite{breuer1} were deposited on Si(111) by Au-induced VLS growth. Au droplets were prepared on the oxide-free substrate by deposition of a 0.6~\AA\ thin Au layer and subsequent heating. Gallium and As fluxes then were set as for sample \#2, and NWs were grown for 30~min at 500$^\circ$C.

For the structural characterization of the NWs, we performed x-ray-diffraction (XRD) and Raman measurements. High-resolution XRD experiments were carried out using a Panalytical X-Pert PRO MRD\texttrademark\ system with a Ge(220) hybrid monochromator and a three bounce Ge(220) channel cut analyzer crystal. We used CuK$\alpha_1$ radiation with a wavelength of $\lambda=1.54056$~\AA. 
 
Continuous-wave micro-photoluminescence ($\mu$PL) and micro-Raman measurements were performed for as-grown NWs and for NWs dispersed on a Si substrate at a temperature of 10~K using the 633~nm line of a He-Ne laser and the 413~nm line of a Kr$^+$ ion laser for excitation, respectively. The laser beam was focused by a 50$\times$ microscope objective to a spot of about 1~$\mu$m diameter. The PL or Raman signal was collected by the same objective, dispersed spectrally in a single spectrograph (600~mm$^{-1}$ grating, 800~mm focal length), and detected by a liquid-nitrogen-cooled charge-coupled device array. The elastically backscattered stray light was suppressed by a notch filter. The Raman signal was not analyzed for its polarization. Due to a NW area density of about 1~$\mu$m$^{-2}$ for all samples, approximately 1--3 NWs were excited by the laser spot while illuminating the as-grown samples from the top during PL measurements. Raman spectra were recorded in backscattering geometry with the light collected perpendicular to the NW axes. 

Spatially resolved measurements of the luminescence spectra along the axis of NWs, which were dispersed on a Si substrate, have been performed by cathodoluminescence spectroscopy (CL) in a scanning electron microscope (SEM). Moreover, secondary electron (SE) and monochromatic CL images were acquired simultaneously for an accurate assignment of the local origin of the CL. The CL/SE experiments were performed using a Zeiss \mbox{ULTRA55} field-emission SEM equipped with a Gatan \mbox{monoCL3} and He-cooling stage system for sample temperatures between 6~K and 300~K. The CL system can be operated with either a photomultiplier tube for monochromatic imaging or a charge-coupled device (CCD) detector to record CL spectra. For the acquisition of both CL spectra and CL images, the acceleration voltage of the electron beam was chosen between 3 and 5~kV, while the beam current was chosen to be 0.5 and 5~nA at 6 and 300~K, respectively. Throughout the CL experiments, the spectral resolution amounted to 0.3~nm, corresponding to a slit width of the CL spectrometer of 0.1~mm.

\section{Results and discussion}

\subsection{Geometry and structure}

\begin{figure}[t]
\centerline{\includegraphics*[width=7cm]{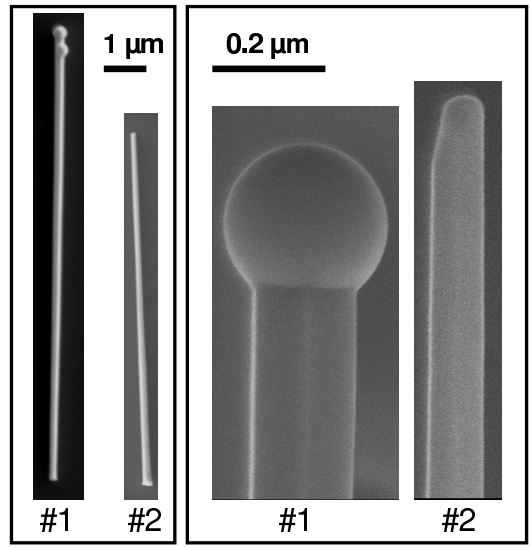}}
\caption{\label{fig1}SEM images of single NWs from sample \#1 and \#2. On the right hand side, the magnified top part of corresponding NWs is shown. }
\end{figure}

Figure \ref{fig1} shows SEM images of single NWs from samples \#1 and \#2 dispersed on a Si substrate. From low- and high-magnification images --- as shown on the left- and right-hand side of Fig. \ref{fig1} --- of a representative number of NWs of both samples, we estimate an average length of the NWs on the order of 9--10 $\mu$m. Their average diameter amounts to 150 and 80~nm for samples \#1 and \#2, respectively. The spheres visible on top of sample \#1 are pure Ga droplets as confirmed by energy dispersive x-ray analysis. The core diameters were estimated assuming negligible axial lengthening during shell growth and considering that the GaAs volumes deposited for core and shell were equal. We, thus, calculate the NW core diameters of samples \#1 and \#2 to be approximately 110 and 50~nm, respectively.

\begin{figure}[t]
\centerline{\includegraphics*[width=8cm]{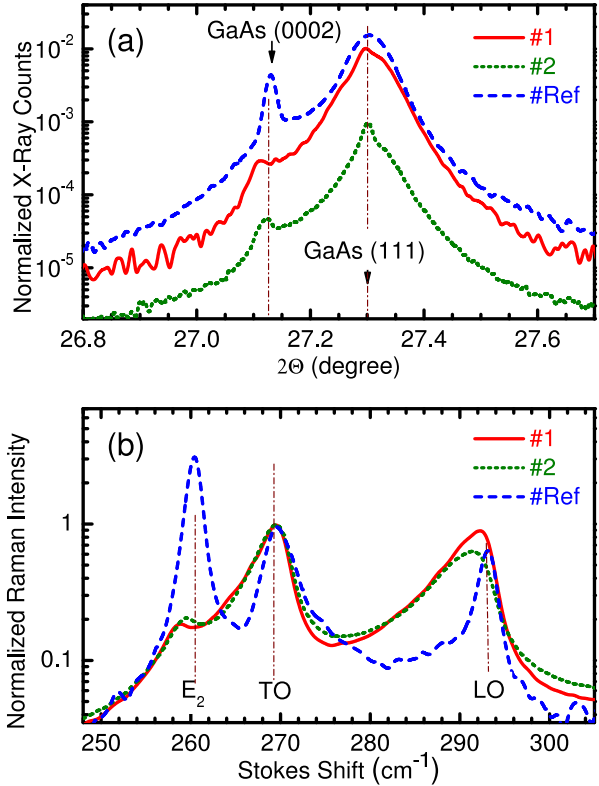}}
\caption{\label{fig2}(Color online) (a) $\omega$-2$\theta$ scans and (b) Raman spectra of the as-grown samples \#1 and \#2 as well as of the reference sample. In (b), $E_2$ and TO (LO) indicate the corresponding phonon lines of GaAs where the $E_2$ phonon mode exists only in the WZ phase.}
\end{figure}

The XRD data of all samples are shown in Fig.~\ref{fig2}(a), which depicts the corresponding $\omega$-2$\theta$ scans. The reflection intensities are normalized to the one of the Si (111) reflection centered at 28.44$^\circ$ (not shown). The observed maxima centered at 27.13$^\circ$ and 27.30$^\circ$ correspond to lattice plane distances of 0.3288 and 0.3264~nm and, therefore, can be assigned to the WZ GaAs (0002) and ZB GaAs (111) reflection, respectively. This assignment is consistent with the assumption that the NWs of the reference sample, for which the narrow peak at 27.13$^\circ$ is most pronounced, crystallize predominantly in the WZ structure. The large width of the GaAs (111) reflection is caused by the fact that parasitic growth leads to an evolution of a highly defective and probably strained GaAs layer in between the NWs, which in addition to the NWs contributes to the ZB reflection. Most importantly, both samples \#1 and \#2 exhibit a weak but distinct peak near the WZ GaAs (0002) reflection indicating that the NWs of both samples contain segments of WZ GaAs. The small shift toward lower angles with regard to the GaAs (0002) reflection of the reference sample is probably caused by the presence of the (Al,Ga)As shell, which is not existent in the reference sample. 

This result is confirmed in Fig.~\ref{fig2}(b), where Raman spectra of the samples \#1 and \#2 as well as of the reference sample are depicted. The Raman intensities are normalized to the peak intensity of the transversal optical (TO) mode of ZB GaAs centered at about 270~cm$^{-1}$. The $E_{2}$ phonon mode, which is most pronounced for the reference sample is characteristic only for WZ GaAs.\cite{zardo} It is also present for samples \#1 and \#2, indicating that the NWs of both samples contain segments of WZ GaAs in accordance with the XRD result of Fig.~\ref{fig2}(a). The shift of the $E_{2}$ phonon line with regard to the reference sample as well as the broadening of the TO and longitudinal optical (LO) resonances toward smaller wave numbers can be explained again by the existence of the (Al,Ga)As shell for samples \#1 and \#2. Summarizing, we emphasize the structural similarity of samples \#1 and \#2 containing comparable portions of WZ GaAs.

\subsection{$\mu$PL of individual GaAs NWs}

\begin{figure}[t]
\centerline{\includegraphics*[width=7cm]{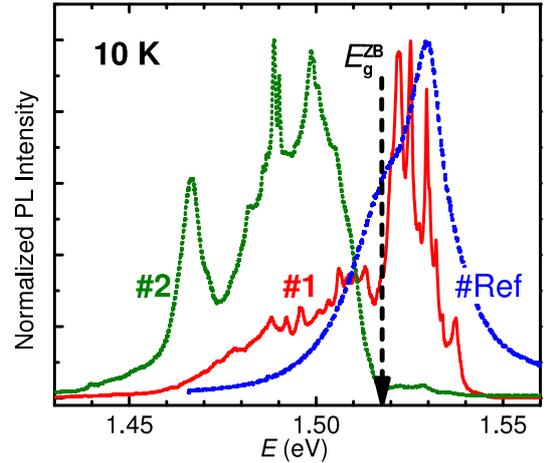}}
\caption{\label{fig3}(Color online) $\mu$PL spectra of as-grown NWs from samples \#1 and \#2 as well as of the reference sample measured at 10 K. $E_g^\mathrm{ZB}$ marks the band gap energy of zincblende GaAs.}

\end{figure}

\begin{figure}[t]
\centerline{\includegraphics*[width=8cm]{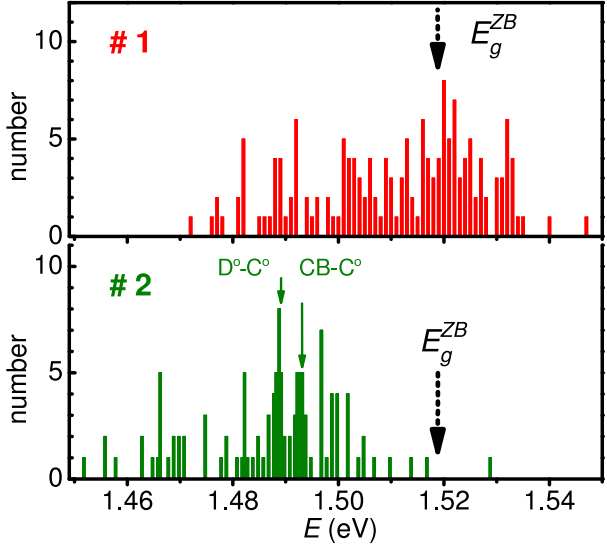}}
\caption{\label{fig4}(Color online) Histogram of peaks found in the $\mu$PL spectra of 30 NWs from samples \#1 and \#2 dispersed on Si(111) as a function of the respective spectral position. $E_g^\mathrm{ZB}$ marks the band gap energy of zincblende GaAs. D$^o$-C$^o$ and CB-C$^o$ mark the energy position of the donor-acceptor pair and conduction-band-acceptor transition with carbon acting as acceptor. The corresponding $\mu$PL spectra have been measured at 10~K.}

\end{figure}

Figure~\ref{fig3} represents the $\mu$PL spectra of the as-grown samples \#1 and \#2 as well as of the reference sample \#Ref. For the low excitation density used to excite the NWs from samples \#1 and \#2, the spectra consist of narrow lines, which appear unchanged for repeated measurements at the same position on the sample. The spectrum of the reference sample, for which an about 4 orders of magnitude higher excitation density was necessary to obtain a PL intensity comparable to that of samples \#1 and \#2,\cite{breuer} is broad and does not show any spike-like structure. Its peak energy exceeds the band gap energy of ZB GaAs ($E_g^\mathrm{ZB}=1.519$~eV)\cite{grilli} by more than 10~meV. The $\mu$PL spectra of samples \#1 and \#2 differ from each other in the spectral distribution of the sharp lines. While for sample \#1 the main part of the spectrum is found above $E_g^\mathrm{ZB}$ (similar to the reference sample), it is found below this energy for sample \#2. To confirm the observation concerning samples \#1 and \#2, we analyzed the $\mu$PL spectra of 30 NWs of each of these samples dispersed on Si(111) and plotted the number of narrow lines observed as a function of the respective spectral position in Fig.~\ref{fig4}. For sample \#1, an essential portion of the narrow lines of the $\mu$PL spectra clearly appears within the spectral range above 1.519~eV, and the highest observed emission energy amounts to 1.547~eV. This observation is in disagreement with the experimental results reported in Refs. \onlinecite{spirkoska} and \onlinecite{heiss}, where GaAs NWs with WZ structure do not show luminescence for energies exceeding $E_g^\mathrm{ZB}$. For sample \#2, almost all the sharp $\mu$PL lines appear below $E_g^\mathrm{ZB}$. We also note that, especially for sample \#2 (cf. bottom part of Fig.~\ref{fig4}), a considerable number of sharp PL lines pile up at energy values matching the transition energies of defect-related PL, namely the donor-acceptor pair and conduction-band-acceptor transition related to carbon as an acceptor.\cite{pavesi}

\begin{figure}[b]
\centerline{\label{fig5}\includegraphics*[width=8cm]{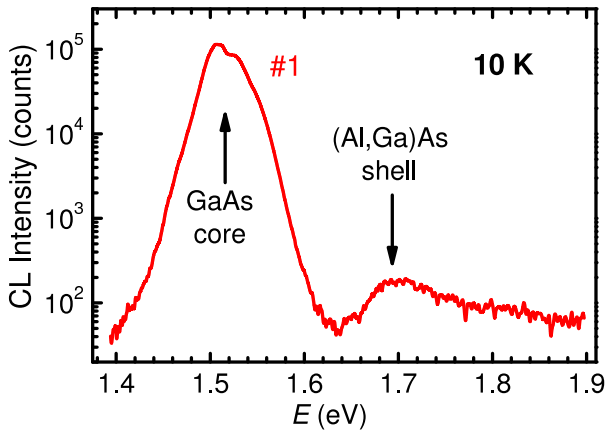}}
\caption{(Color online) CL spectrum of a set of NWs from sample \#1 measured at 10~K.}

\end{figure}

Summarizing, we can establish that although the NWs of both samples \#1 and \#2 contain segments of WZ GaAs with similar volume fraction, their optical emission characteristics differ substantially from each other. While the spectral distribution of sample \#1 resembles the one of the reference sample, for which the main part of the spectrum appears above $E_g^\mathrm{ZB}$, the spectra of sample \#2 do not exceed  $E_g^\mathrm{ZB}$ as observed by other authors \cite{spirkoska,heiss}.

\begin{figure}[t]
\centerline{\includegraphics*[width=7cm]{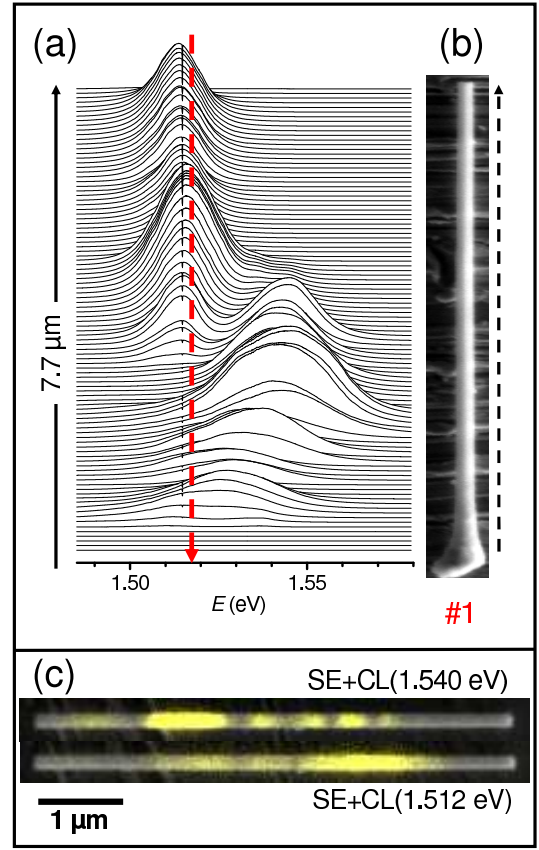}}
\caption{\label{fig6}(Color online) (a) CL spectral line scan along the axis of a single NW from sample \#1. The dashed vertical arrow marks the band gap energy of ZB GaAs. (b) SEM image of the corresponding NW, where the dashed arrow marks the path of the scan. (c) SE image of another NW from sample \#1 superimposed by CL images obtained for a detection energy of 1.540~eV (top) and 1.512~eV (bottom). Both CL spectra and images have been acquired at 6~K.}

\end{figure}

At this point, we would like to comment on the lateral confinement of carriers in one-dimensional structures as a possible reason for a blue-shift of the corresponding transition energies.
For our NWs, such a quantum effect can be ruled out, since the estimated values of the NW core diameters are too large to cause a confinement-related shift of the transition energy above $E_g^\mathrm{ZB}$. Moreover, we observe $\mu$PL lines exceeding $E_g^\mathrm{ZB}$ for the NWs with a larger diameter (\#1) and not for those with a smaller diameter (\#2).  

\subsection{CL of individual GaAs NWs}

 \begin{figure}[t]
\centerline{\includegraphics*[width=7cm]{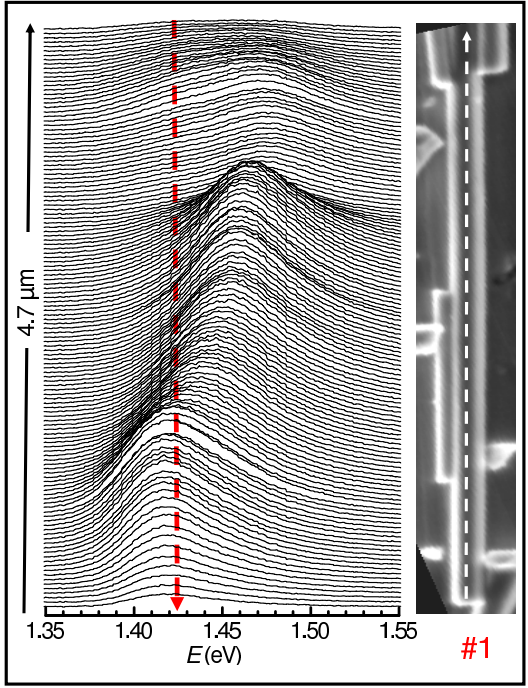}}
\caption{\label{fig7}(Color online) CL line scan along the axis of a single NW from sample \#1 acquired at 300~K. The dashed vertical arrow marks the band gap energy of ZB GaAs at 300~K. The inset shows an SEM image of the corresponding NW, where the dashed arrow marks the path of the scan.}

\end{figure}

In order to further elucidate the luminescence distribution of the NWs from sample \#1, we performed spatially resolved CL measurements on single NWs at 6~K and 300~K.
In Fig.~\ref{fig5}, we show a low-temperature CL spectrum acquired from a set of dispersed NWs of sample \#1 over a wide spectral range. This spectrum consists of a broad intense luminescence band centered at 1.51~eV and a three orders of magnitude weaker band centered at 1.70~eV. We assign the former, which corresponds to the $\mu$PL emission in Figs.~\ref{fig3} and ~\ref{fig4}, to the GaAs core and the latter to the (Al,Ga)As shell. The assignment to the shell emission is suggested by its peak position, which corresponds to an Al content of 12\% in reasonable agreement to the nominal value of 10\%. Moreover, the low CL intensity of the high-energy band is consistent with an efficient loss of carriers excited in the shell due to both carrier capture within the GaAs core and non radiative recombination at the shell surface.

\begin{figure}[t]
\centerline{\includegraphics*[width=8cm]{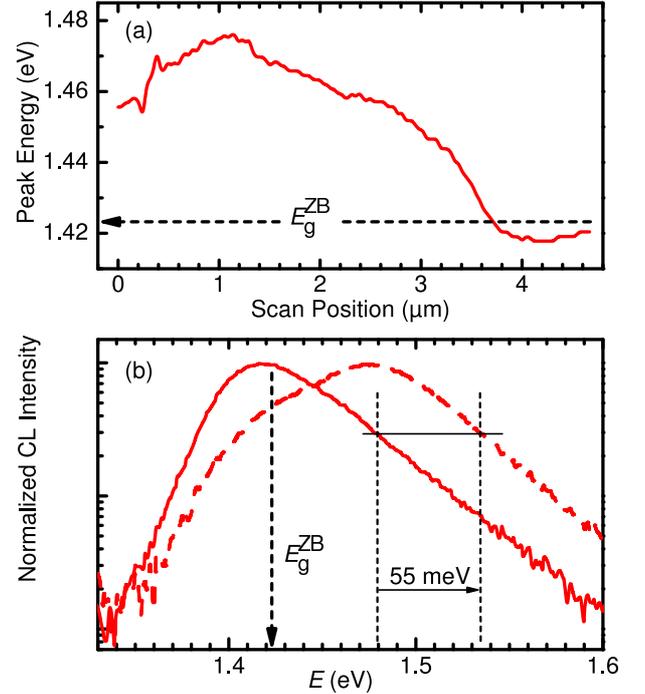}}
\caption{\label{fig8}(Color online) (a) Peak energy as a function of scan position of the CL spectra in Fig.~\ref{fig7}. (b) CL spectra showing the lowest and highest peak position among the spectra of Fig.~\ref{fig7}. The dashed arrows mark the band gap energy of ZB GaAs at 300~K. At the high energy side of the spectra, the exponential decay reflects the thermal distribution of the carriers at the band edge. The spectral shift of this exponential decay when measured at different positions along the NW axis is assigned to the band gap difference $E_g^\mathrm{WZ}-E_g^\mathrm{ZB}$.}

\end{figure}

\begin{figure*}[t!]
\centerline{\includegraphics*[width=12cm]{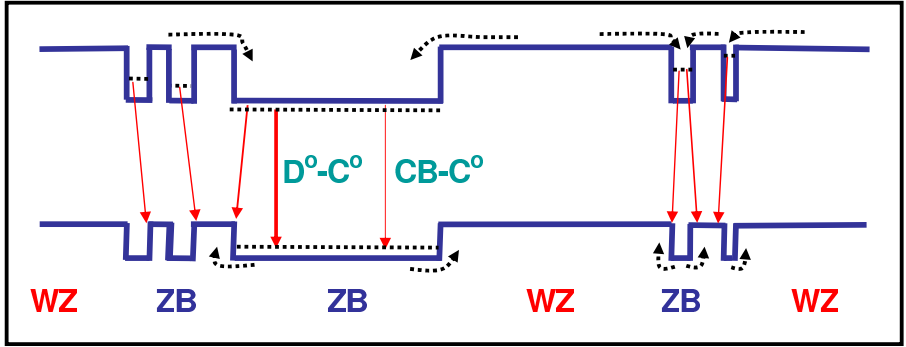}}
\caption{\label{fig9}(Color online) Qualitative model of the band gap alignment for samples \#1 and \#2. ZB and WZ mark zincblende and wurtzite GaAs segments, respectively. CB-C$^\circ$ and D$^\circ$-C$^\circ$ mark conduction-band-acceptor (carbon) and donor-acceptor (carbon) transitions, respectively.}

\end{figure*}

Figure~\ref{fig6} visualizes the spatial and spectral distribution of the GaAs core emission along the axis of individual NWs from sample \#1, where the CL spectra shown in Fig.~\ref{fig6}(a) were obtained by scanning the electron beam along the axis of the NW depicted in Fig.~\ref{fig6}(b). Clearly, the spectral contributions centered near and significantly above $E_g^\mathrm{ZB}$ are spatially separated from each other along the NW axis. While the upper part of the NW under investigation emits light predominantly near the band gap of ZB GaAs, the CL spectra of the lower NW section are found between $E_g^\mathrm{ZB}$ and 1.55 eV. In Fig.~\ref{fig6}(c), we show SE images of another NW from the same sample superimposed by monochromatic CL images acquired for CL detection energies below and above $E_g^\mathrm{ZB}$. The CL obtained at 1.540~eV is found within well separated segments. Due to carrier diffusion along the NW axis, the segment structure of the CL image is smeared out for the image obtained at the lower detection energy (1.512~eV). 
These observations demonstrate that the WZ part of the NWs detected by XRD and Raman (cf. Fig.~\ref{fig2}) is indeed distributed segment-like along the NW axis and that $E_g^\mathrm{WZ} > E_g^\mathrm{ZB}$. 

Also for sample \#2, we measured the variation of the CL spectra along the axis of individual NWs (not shown here). Again, the spectral position of the CL varies along the NW axis within a wide range. However, for a large portion of each individual NW, the emission is found at the spectral positions assigned to carbon-related defects, in agreement with the $\mu$PL data of Fig.~\ref{fig4}. In contrast to sample \#1, only negligibly small fractions of the NWs from sample \#2 show CL at energies higher than $E_g^\mathrm{ZB}$.

We expect that the energy of the near-band-edge luminescence reflects the actual value of the band gap when measuring at 300~K rather than at low temperatures, since shallow defects are ionized at higher temperatures. Thus, to obtain a quantitative estimate of the actual difference $E_g^\mathrm{WZ}-E_g^\mathrm{ZB}$, we performed CL spectral line scans along single NWs from sample \#1 at 300~K.
The corresponding CL spectra acquired along the axis of a representative NW are shown in Fig.~\ref{fig7}. Again, one section of the NW emits at energies centered at $E_g^\mathrm{ZB}$, but others emit light at energies significantly exceeding $E_g^\mathrm{ZB}$. The variation of the emission energy along the axis of the NW in Fig.~\ref{fig7} is plotted in Fig.~\ref{fig8}(a). According to this graph, the maximum energy variation along this NW amounts to 58~meV. 
For a more accurate estimate of the variation of the band gap energy along the NW, in Fig.~\ref{fig8}(b), we compare the spectra with the highest and lowest peak energy. Note that the CL intensities are represented on a logarithmic scale. The thermal distribution of carriers is reflected by an exponential tail of the CL intensity on the high-energy side of the spectra. The slope of this tail is the same for all spectra shown in Fig.~\ref{fig7} and indicates the actual temperature of the free carriers under the chosen excitation conditions. Thus, the energy shift of the exponential tail between the spectra of Fig.~\ref{fig8}(b) represents the maximal energy difference that we have measured along the NW. We assume this difference amounting to 55~meV to be the lower bound for the band gap difference between WZ and ZB GaAs.

\subsection{Discussion and qualitative model}

Our experimental results can be understood within the framework of a quantum well (QW) model for heterostructures consisting of WZ and ZB GaAs as has been recently reported also for GaN nanowire WZ/ZB heterostructures.\cite{jacopin} According to the calculations of Murayama and Nakayama \cite{murayama} for GaAs, such a heterostructure establishes a type II band alignment with positive WZ conduction and valence band offsets. Thus for NWs consisting of alternate ZB and WZ GaAs segments, electrons captured within the ZB segments recombine with holes of the neighboring WZ segments as illustrated in Fig.~\ref{fig9}. For low $T$, this results in excitons bound to the ZB/WZ interface. The corresponding transition energy $E_t$ depends on the degree of confinement of the electrons and can span the range between $E_g^\mathrm{WZ}$ and $E_g^\mathrm{WZ}-\Delta E_C$, where $\Delta E_C$ is the conduction band offset. Values for $E_t$ exceeding $E_g^\mathrm{ZB}$ as observed for sample \#1 are possible for very thin ZB segments as a result of confinement. For simplicity, we ignore the possibility of an additional confinement of holes in thin WZ segments of a more complicated heterostructure as we probably find it in our NWs. This would, however, be part of a sequence of closely spaced ZB and WZ segments where electronic coupling of the QWs would counteract the additional confinement. 

\begin{figure}[b]
\centerline{\includegraphics*[width=8cm]{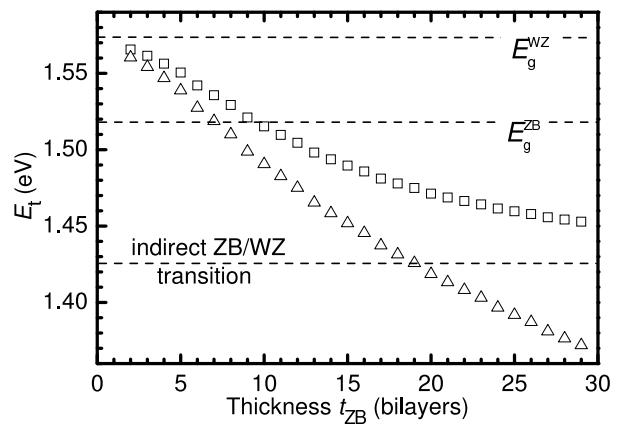}}
\caption{\label{fig10}Calculated transition energy of a WZ/ZB/WZ GaAs QW structure as a function of the ZB segment thickness, where the confined electron of the ZB layer recombines with a hole of the neighboring WZ GaAs. The energy is plotted both for the case of $P_\mathrm{sp}=0$ (squares) and $P_\mathrm{sp}=0.002$~Cm$^{-2}$ (triangles).}

\end{figure}

In order to obtain an estimate of $E_t$ as a function of the thickness of the ZB GaAs segments ($t_\mathrm{ZB}$), we have performed selfconsistent Poisson-Schr\"odinger calculations\cite{1DPoisson} for low $T$ giving the ground state of electrons in a WZ/ZB/WZ GaAs quantum well (QW) structure, where the confined electron of the ZB layer recombines with a hole of the neighboring WZ GaAs. We assume a WZ conduction band offset with respect to ZB GaAs of $\Delta E_C=+159$~meV \cite{bechstedt}, and we take into account a band gap difference of 55~meV, which we found experimentally to be the lower bound of the difference $E_g^\mathrm{WZ}-E_g^\mathrm{ZB}$. Standard values for the effective masses are employed. The results of our calculations are summarized in Fig.~\ref{fig10}, where $E_t$ is shown as a function of $t_\mathrm{ZB}$. The theoretical data of Fig.~\ref{fig10} predict values for $E_t$ ranging between $E_g^\mathrm{WZ}$ ($\approx 1.57$~eV) and 1.43~eV when $t_\mathrm{ZB}$ varies between 3 and 30 bilayers (a bilayer being one layer of Ga and As amounting to $t=c/2$ with $c$ being the WZ $c$-lattice constant).
It becomes evident that emission energies higher than $E_g^\mathrm{ZB}$ are expected only for very thin ZB insertions, i.e., $t_\mathrm{ZB}<9$~bilayers. 

A spontaneous polarization $P_\mathrm{sp}$ along the (0001) direction is a property intrinsic to WZ semiconductors but vanishes for ZB phases due to their higher symmetry. As the latter is the equilibrium phase for GaAs and an interest in the WZ polytype has only recently come up with the advent of NW growth, the possibility of a spontaneous polarization in WZ GaAs has not been treated in the literature. For the investigation of the impact of $P_\mathrm{sp}$ on of $E_t(t_\mathrm{ZB})$, the calculations have further been performed taking a spontaneous polarization into account. The deviation from the ideal tetrahedral coordination of the atoms characterized by the internal parameter $u$ is comparable---but opposite in direction---to that of GaN\cite{mcmahon}, while the bond ionicity is lower than that of GaN. Thus, we assume that the spontaneous polarization of GaAs is opposite in sign and about one order of magnitude lower than that of GaN, and we estimate $P_\mathrm{sp}=0.002$~Cm$^{-2}$ \cite{laehnemann}. As shown in Fig.~\ref{fig10}, this small value of $P_\mathrm{sp}$ leads to a significant increase of the slope of $E_t(t_\mathrm{ZB}$) and, consequently, to a larger reduction of $E_t$ with increasing thickness of the ZB insertion as for $P_\mathrm{sp}=0$. Accordingly, energies higher than $E_g^\mathrm{ZB}$ should be observable only for $t_\mathrm{ZB}<7$~bilayers. 

In the framework of the above model, the different emission properties of samples \#1 and \#2 can be explained by assuming a higher probability for the occurrence of very thin ZB sections ($t_\mathrm{ZB}<7$~bilayers) for sample \#1, whereas for sample \#2 the ZB sections are in general thicker than 7--9~bilayers. While for the former, the electron confinement leads to values of $E_t$ exceeding $E_g^\mathrm{ZB}$, the electron confinement is small for sample \#2 and $E_t$ is generally found below $E_g^\mathrm{ZB}$.

Zincblende insertions of two, three, and four bilayers can be considered as different kinds of stacking faults in WZ GaAs. As the corresponding QW interfaces are atomically smooth, such stacking fault QWs of different thicknesses should result in distinct sharp luminescence lines. This phenomenon of luminescence from excitons bound to stacking faults is actually well known for WZ GaN micro structures\cite{rebane,sun,skromme,laehnemann}. Indeed, the $\mu$PL spectra shown in Fig.~\ref{fig3} exhibit distinct and sharp emission lines. A direct comparison of the energy of these sharp lines with the calculated $E_t(t_\mathrm{ZB})$ of Fig.~\ref{fig10}, however, is not conclusive because of the simplifications made in the calculation.

\section{Conclusions}

Luminescence of GaAs nanowires consisting of wurtzite and zincblende segments is spread over a wide spectral range, which we explain in terms of a type II band alignment of the ZB/WZ heterostructure sequence.  Depending on the growth conditions, we observe luminescence below and clearly above the band gap energy of ZB GaAs. The latter indicates a wider band gap for the WZ as compared to the ZB modification of GaAs. The thickness of the ZB insertion controls the energy of the transition between electrons confined in this insertion and holes of the neighboring WZ sections. This energy can exceed $E_g^\mathrm{ZB}$ for ZB insertions as thin as a few bilayers as a result of the confinement of the electrons. Obviously, this condition is fulfilled for sample \#1 investigated in this study, while for sample \#2 the thickness of the ZB segments of the NWs must be generally larger than the above mentioned critical value, even though the total amount of ZB and WZ in the two samples is comparable according to the results from XRD and Raman measurements. Thus, our results clarify why in some previous works luminescence was not observed at energies above $E_g^\mathrm{ZB}$ \cite{spirkoska,moewe,novikov,heiss,ketterer,ketterer1}, while in other works luminescence was also observed above the band gap energy of ZB GaAs\cite{jancu,hoang1,hoang2,ihn}.

At low temperatures, defect-related optical transitions contribute essentially to the observed diversity and high probability of luminescence lines below $E_g^\mathrm{ZB}$. In particular for sample \#2, an accumulation of emission lines at energies, which can be assigned to the donor-acceptor-pair and conduction-band-acceptor transition, have been observed, where carbon acts as an acceptor. Acceptors exhibit a large carrier capture cross section and can, thus, effectively bind holes within the ZB segments, namely preventing the hole diffusion toward the WZ segments. Thus, an absence of emission lines above $E_g^\mathrm{ZB}$ is very probable especially for low temperatures, but does not mean an absence of higher energy states. 
The possible energy range of optical transitions increases, if spontaneous polarization is taken into account. 

By means of spatially resolved CL measurements at 300~K, we estimate the lower bound for the difference $E_g^\mathrm{WZ}-E_g^\mathrm{ZB}$ to be 55~meV.

\section{Acknowledgement}

We gratefully appreciate the assistance of David Stowe, Gatan Inc., for the acquisition of the CL line scan represented in Fig.~\ref{fig7}.
We thank Rudolf Hey and Holger T. Grahn for a critical reading of the manuscript.

\end{document}